\documentclass[paper]{JHEP3} % 10pt is ignored!

\usepackage{epsfig,multicol,bbm}

\newcommand\fverb{\setbox\pippobox=\hbox\bgroup\verb}
\newcommand\fverbdo{\egroup\medskip\noindent%
			\fbox{\unhbox\pippobox}\ }
\newcommand\fverbit{\egroup\item[\fbox{\unhbox\pippobox}]}
\newbox\pippobox
\title{ About the helix structure of the Lund string }

\author{\v{S}\'{a}rka Todorova-Nov\'a \\ Tufts University \\
	E-mail: \email{sarka.todorova@cern.ch}
}

\abstract{ The helix structure of the Lund string,
 first derived from studies devoted to the emission of soft gluons at the end
 of the parton cascade, may be at the origin of certain characteristic discrepancies observed 
 in the low transverse momentum region at LEP and LHC. A study of the relation between different
 helix implementations and observable effects is presented. The model is extended to cover
 a multiparton string topology (result of parton shower), and compared with the experimental data.
 It is found that a helix-ordered string with a regular winding (proportional to the energy density stored in the string),
 is favoured by the inclusive single-particle spectra measured in the hadronic decay of Z$^0$.  
}


\begin{document} 

%\maketitle  IS IGNORED %%%%%%%%%%%

\section{Introduction}

   The Lund fragmentation model \cite{lund,Sjostrand:2006za}
 uses the concept of string with uniform energy density to model the confining colour field between partons carrying
 complementary colour charge. The string is viewed as being composed of straight
 pieces stretched between individual partons according to the colour flow.
 The fragmentation of the string proceeds via the tunneling effect (creation
 of a pair of a quark and an antiquark from vacuum) with a probability
 given by the fragmentation function. The sequence of string break-ups defines the
 final set of hadrons, each built from a $q\bar{q}$ pair (ev.diquark in case
 of baryons) and a piece of string between the two neighbouring string break-ups.
 The longitudinal hadron momenta stem directly from the space-time
 difference between the break-ups.
   
 The model gives a fair description of the available high-energy hadronic
 data and is therefore widely used in experimental particle physics.  
 It reproduces particularly well the particle multiplicity and
 longitudinal profile (jet formation) but there are certain characteristic
 discrepancies between data and simulation which suggest the treatment of transverse momenta 
 may not be entirely adequate ( more in Section 6). It is therefore
 interesting to develop and study alternative models.
 
  A very interesting work devoted to the study of the properties
 of the emission of soft gluons was published by Andersson et
 al. some time ago \cite{lund_helixm}. Under the assumption that the generating current has a tendency to emit as many
 soft gluons as possible, and due to the  constraint imposed on the emission angle
 by helicity conservation, it was shown that the optimal packing of emitted gluons in the phase space
 corresponds to a helix-like ordered gluon chain. Such a structure of the colour field cannot
 be expressed through gluonic excitations of the string and it needs to be implemented as an internal string property.

\section{Fragmentation of the Lund string with helix structure} 

    The implementation of a string with a helix structure radically
 changes the way hadrons acquire their transverse momentum.
 In the conventional Lund model \cite{lund}, the transverse momentum
 of the hadron is the (vectorial) sum of the transverse momenta of the
 (di)quarks which were created via a tunneling process during the
 breakup of the string. The transverse momenta of newly created partons
 are randomly sampled from a gaussian distribution (with adjustable width) and their
 azimuthal direction is random.

    In the helix ordered string, hadrons obtain their transverse momentum from the shape of the colour
 field itself, so that there is in principle no need to assign a momentum to new quarks in the string
 breakup. If we picture the colour field as a stream of soft gluons ordered at emission,
 we get the hadron transverse momentum by integration over the transverse momenta
 of soft gluons emitted in between the string break-up points which define the hadron, see Fig.\ref{fig:helix}:

     \[\vec{p}_{\rm t}(hadron) = \int_{\Phi_i}^{\Phi_j}
 \vec{p}_{\rm t}(gluon){\rm d}\Phi \]
   where $\Phi_{i(j)}$ is the 'phase' of the helix (azimuthal angle)
 in the break-up point $i(j)$.

%\begin{figure}[h]
%\begin{center}
\FIGURE[bh]{
  \mbox{\epsfig{file=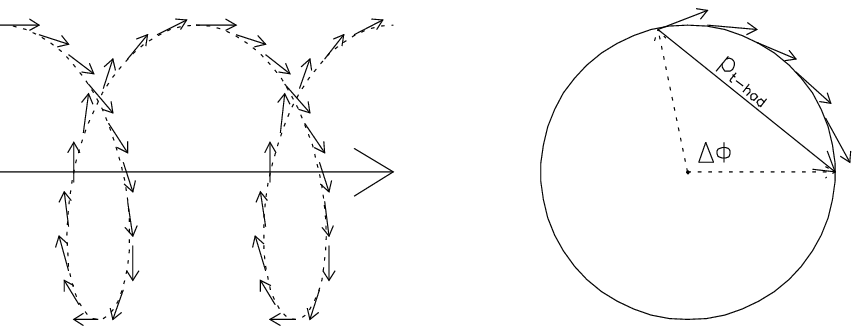,height=6.cm}}
\vspace{-1.cm}
 \hspace{6.cm} a) \hspace{5.cm} b)
 \caption{\protect\small    \sl
 a) The helix structure of the string carried by colour connected chain of soft gluons.
 b) After fragmentation, the transverse
 momentum of a direct hadron is the integral of the transverse momenta
 of the soft gluons, integrated over the corresponding string piece.
\label{fig:helix}}
 }
%\end{center}
%\end{figure}

   The transverse momentum the hadron carries is thus entirely defined by the properties of the helix ordered field.
 This additional constraint translates into a loss of azimuthal degree of freedom in the string break-up,
 arguably the most significant consequence of the implementation of the helix string model in the fragmentation process.
 The underlying helix structure is reflected in correlations between transverse and longitudinal components of hadrons 
 which may lead to experimentally observable effects, depending on the actual form of the helix string.

   So far, only one type of helix string parametrization was put under scrutiny \cite{lund_helixm,alessandro}, and no
 convincing experimental evidence in favour of the helix string was obtained. The purpose of the present paper is to
 introduce an alternative helix string parametrization, to study and compare the observable effects, and to show 
 that the helix string model (after modification) describes the hadronic data better than the standard Lund fragmentation model.

\section{Parametrization of the helix string: theory}

   As a reminder, and for the sake of clarity, we reiterate the properties
 of the helix string introduced in \cite{lund_helixm}. Certain aspects of the original implementation which
 were not necessarily addressed in the original paper, but which are relevant for the discussion, will be pointed out.

\subsection{The Lund helix model}

  In \cite{lund_helixm}, the phase difference of the helix winding was related
 to the rapidity difference of the emitting current by the formulae:
 
\begin{equation}
        \Delta\Phi = \frac{\Delta y}{\tau},
\label{eq:lund_helix}
\end{equation}

 where $\Delta\Phi$ is the difference in helix phase between two points along the string, 
 $\Delta y$ is their rapidity difference, and $\tau$ is a parameter.
 In the Lund model, the rapidity at a given point along the string is defined as 

 \begin{equation}
       y = 0.5 \ {\rm ln}(\frac{k^+  p^+}{k^-  p^-}),
\label{eq:rapidity}
\end{equation}

    where $p^{+,-}$ are the initial light-cone momenta of the 
 endpoint quarks and $k^{+,-}$ their \emph{fractions}
 defining a position along the string, see Fig.\ref{fig:def}.

%\vspace{-1.cm}
%\begin{figure}[h]
%\begin{center}
\FIGURE[h]{
  \mbox{\epsfig{file=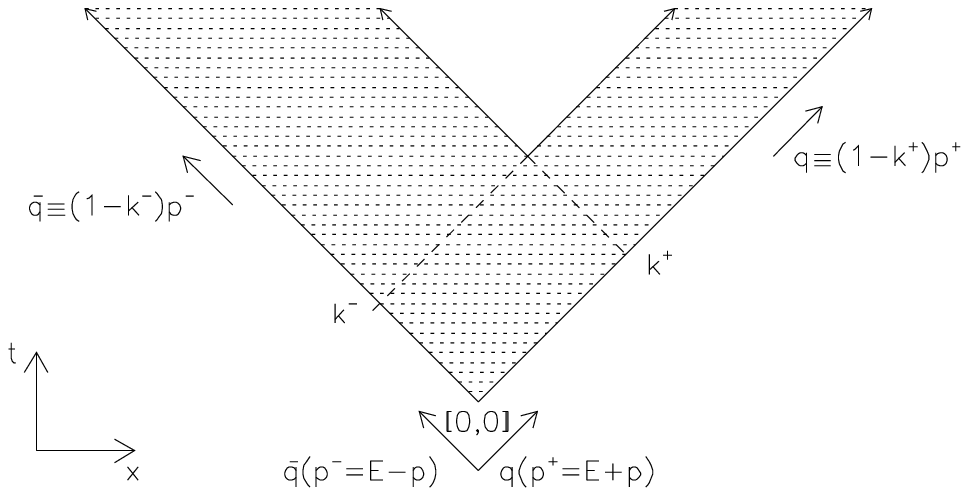}}
 \caption{\protect\small \sl Evolution of the string in the rest frame
 of the $q\bar{q}$ pair, including the first string break-up.
 The partons lose their momentum as they
 separate and the string - the confining field - is created.
 The space-time coordinates
 can be obtained from the relation
 [t,x]=(k$^+$p$^+$+k$^-$p$^-$)/$\kappa$ ($\kappa  \sim$ 1 GeV/fm).
 The x direction is parallel to the string axis.
\label{fig:def}
}
}
%\end{center}
%\end{figure}

   The rapidity difference between two points along the string is then
\begin{equation}
        \Delta y = 0.5 \ {\rm ln}(\frac{k_i^+ k_j^-}{k_i^- k_j^+})
\label{eq:dy}
\end{equation}
   and it is related to the {\it angular} difference  of points 
 in the string diagram (Fig.\ref{fig:def}). 

\FIGURE[hbt]{
  \mbox{\epsfig{file=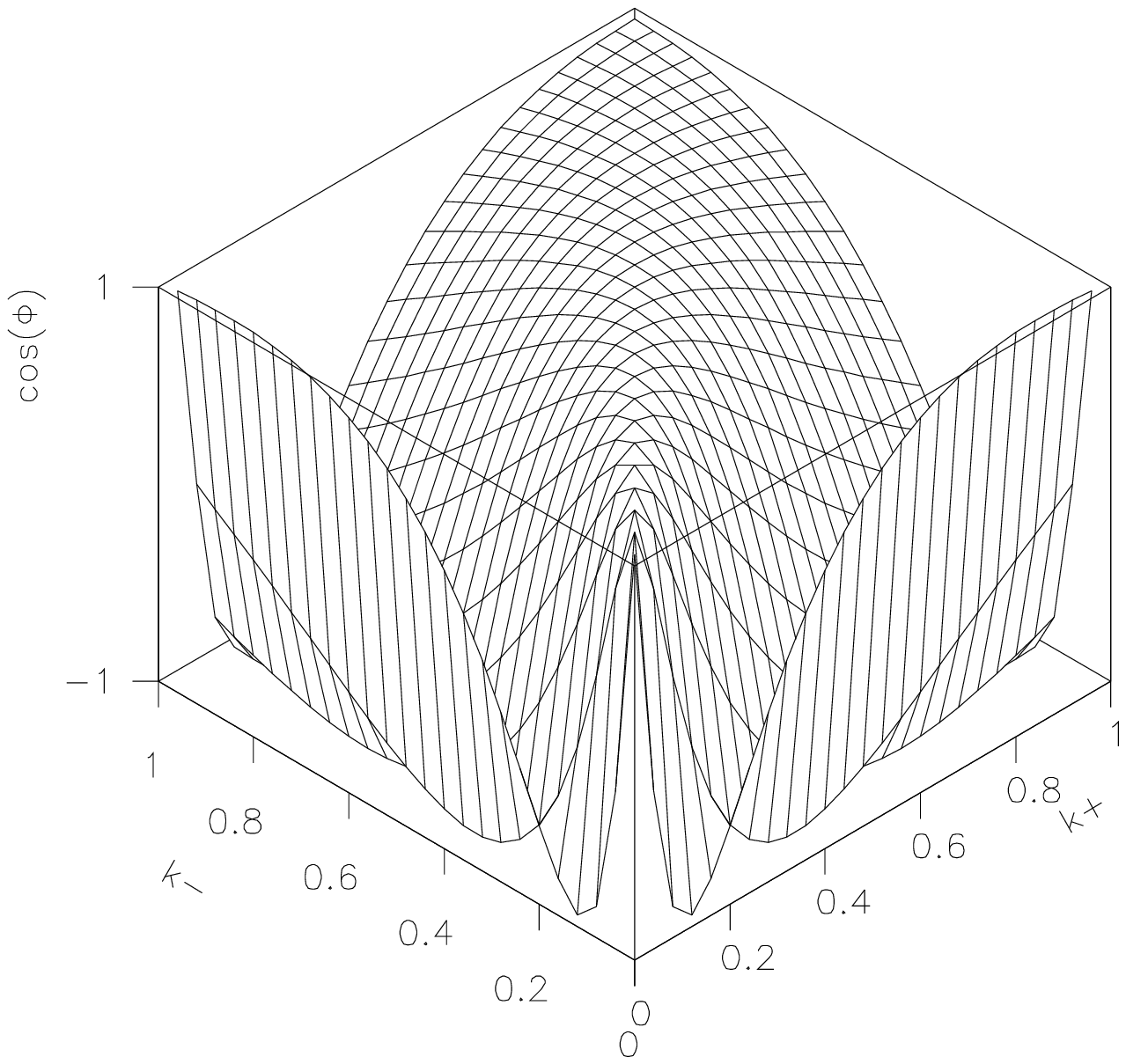,width=10.cm}
        \epsfig{file=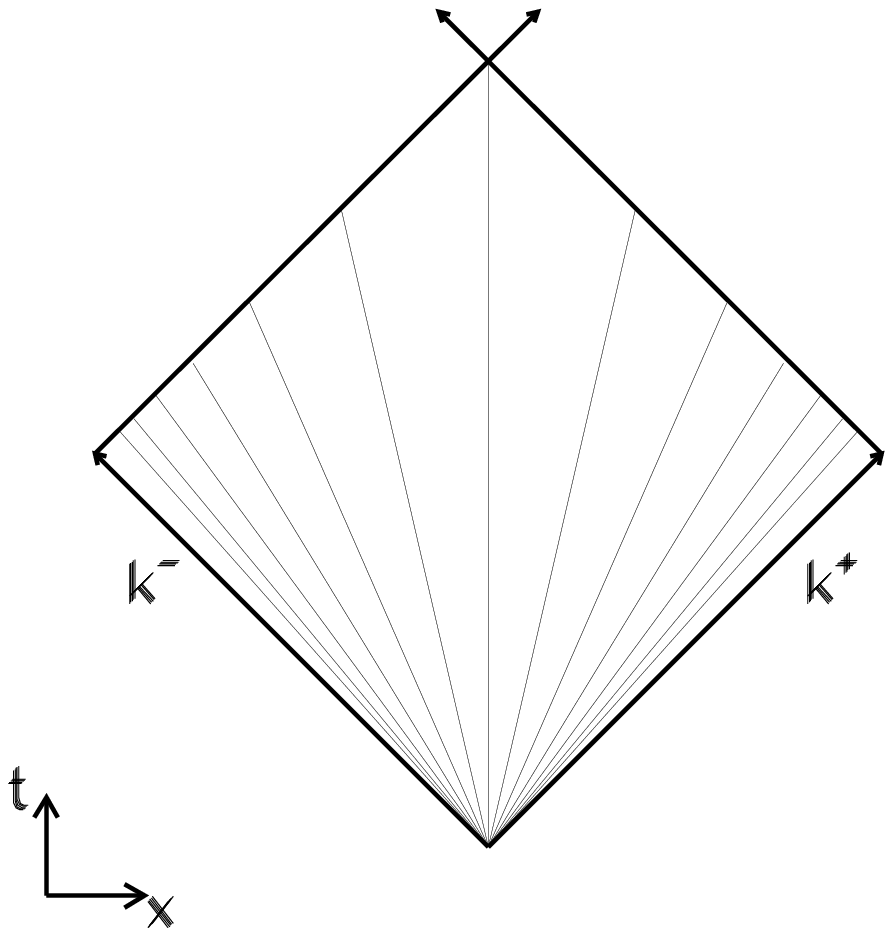,width=5.cm}}
\vspace{-0.5cm}
% \hspace{6.cm} a) \hspace{5.cm} b)
 \caption{\protect\small  \sl
 On the left: the phase of the helix winding along the string
 according to the Lund prescription (Eq.\ref{eq:lund_helix})
 for parameter $\tau$=0.3.
 The phase was fixed by a random choice at one point. On the right:
 for better illustration of the phase evolution in the space-time
 coordinates, the points with equal phases are connected by lines
 in equidistant phase intervales. The singularity present in the model
 at the endpoint of the string is not graphically emphasized.
 \label{fig:lund_phase} }
}

    The evolution of the phase of the helix string defined 
 according to (\ref{eq:lund_helix})
 is illustrated for a simple $q\bar{q}$ string in the
 string diagram Fig.\ref{fig:lund_phase}.
 The phase is fixed by a random choice at one point of the
 diagram (in our example $\Phi$=0 at [k$^+$,k$^-$]=[1,1])
 and is calculated
 for the rest of the diagram from Eq.\ref{eq:lund_helix} with the help of 
 Eq.\ref{eq:rapidity}. The parameter $\tau$ is set to 0.3 for definiteness,
 its value is irrelevant for discussion of the qualitative features of
 the model. 
 Please note that the density of helix winding increases with the distance from the string center, and becomes infinite
 near the turning points ([k$^+$,k$^-$]=[0,1]/[1,0]) where Eq.~\ref{eq:dy} contains a singularity.
 This singularity does not affect the modelling of simple $q\bar{q}$ strings as long as the endpoint quarks
 do not acquire transverse momentum (a default solution in Pythia \cite{Sjostrand:2006za}) but the model needs some sort of regularization
 in case of multiparton string configurations. (All studies in \cite{lund_helixm} were done using a simple $q\bar{q}$ string.)   
   
\subsection{The modified helix model}

   The presence of a singularity in the Lund helix model is one of the reasons why we wish to take a second look
 at the definition of the helix model. Also the requirement of the homogenity of the string field which lies
 at the heart of the Lund fragmentation model seems to be poorly satisfied, given the difference of the
 helix winding at the middle of the string and near the string endpoints. It should be emphasized that
 we strictly adhere to the central idea of \cite{lund_helixm}, namely the emergence of a helix-ordered
 gluon chain at the end of the parton cascade, and that we are merely looking into details of the helix parametrization. 

    We derive the alternative helix model studying the properties of an elementary dipole in the gluon chain
 on the basis of equation 3.3 from \cite{lund_helixm}.
  The squared mass of the dipole formed by colour connected gluons can be written as
   
 \begin{equation}
      s_{j,j+1} = k_T^2 \ 2 \ [cosh(\Delta y) -  cos(\Delta\Phi)] 
 \label{eq:mass} 
 \end{equation}             

   where the transverse momenta of both gluons are set to $k_T$ (for simplicity) ,$\Delta y$ is the
 difference in rapidity, and $\Delta\Phi$ difference in azimuthal angle between gluons.

   The original proposal for the helix string neglected the azimuthal difference
 in the search for gluon packing which would minimize the gluon distance yet satisfy the helicity conservation laws,
 and the distance between gluons was parametrized with the help of their rapidity difference. Here we
 intend to reverse the approach and develop a helix model which minimizes the rapidity difference between
 soft gluons and where the gluons are separated mainly in the transverse plane.
%
% For the discussion
% of the corresponding physics picture see Appendix A.
 
   Under the assumption
\begin{equation}
      \Delta\Phi >> \Delta y \approx 0
\end{equation}
   equation (~\ref{eq:mass}) reads
\begin{equation}      
     s_{j,j+1} = k_T^2 \ 2 \ [ 1 -  cos(\Delta\Phi)] = 4 \  k_T^2 \ sin^2(\frac{\Delta\Phi}{2})
\end{equation}
   and the distance $d$ between gluons, as introduced in \cite{lund_helixm}, becomes
\begin{equation}      
     d_{j,j+1} = \sqrt{ s_{j,j+1}/ k_T^2 } = \ 2 \ | \ sin(\frac{\Delta\Phi}{2}) \ |
\end{equation}

   The condition $ d \geq \frac{11}{6} $, derived from helicity conservation restrictions,
 is satisfied for $\Delta\Phi > 2.3 $ rad. Since there is no constraint on the length of the gluon
 chain ordered in azimuthal angle, the number of soft gluons in the chain will depend on $k_T$
 and the energy available for string build-up. It has to be stressed however that we assume the 
 emergence of the ordered helix field occurs in parallel with the 'homogenization' of the
 string field in which the interactions between field quanta redistribute the longitudinal
 momenta of field creating partons, and that we can describe the string with the help of uniform
 energy density and string tension, much as the standard Lund fragmentation model does.    .

 We set the difference in the helix phase to be proportional to the energy stored in between
 two points along the string

\begin{equation}
     \Delta \Phi = {\cal S} \ (\Delta k^+ + \Delta k^-) \ M_0/2 ,
\label{eq:my_helix}
\end{equation}
  where $M_0$ stands for the invariant mass of the string, $\cal S$[rad/GeV] is
a parameter, and fractions $\Delta k^+= |k^+_j - k^+_{j+1}|$, $\Delta k^-= |k^-_j-k^-_{j+1}|$ define the size of the string piece. 

\FIGURE[h] {
  \mbox{\epsfig{file=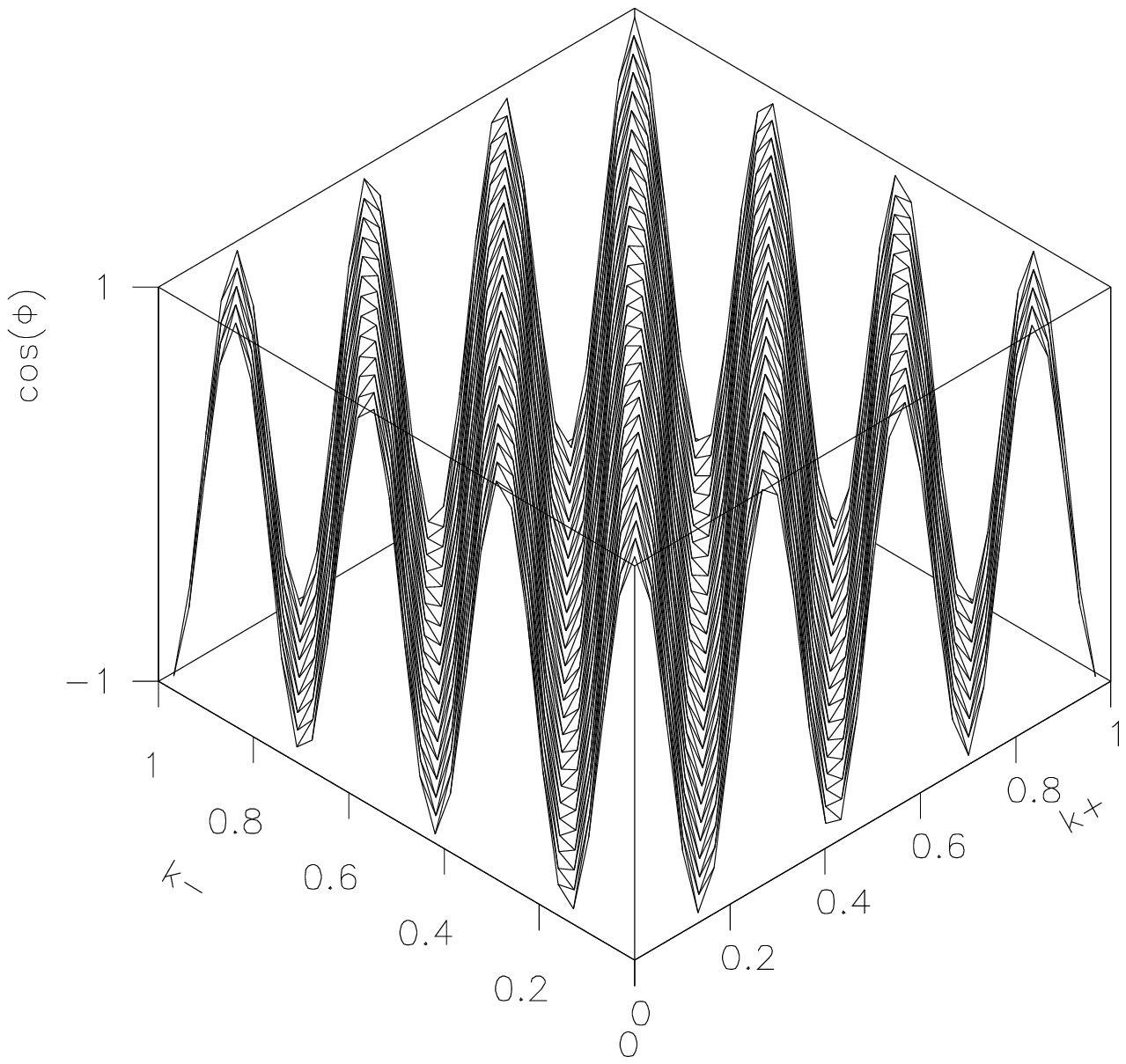,width=9.cm}}
  \mbox{\epsfig{file=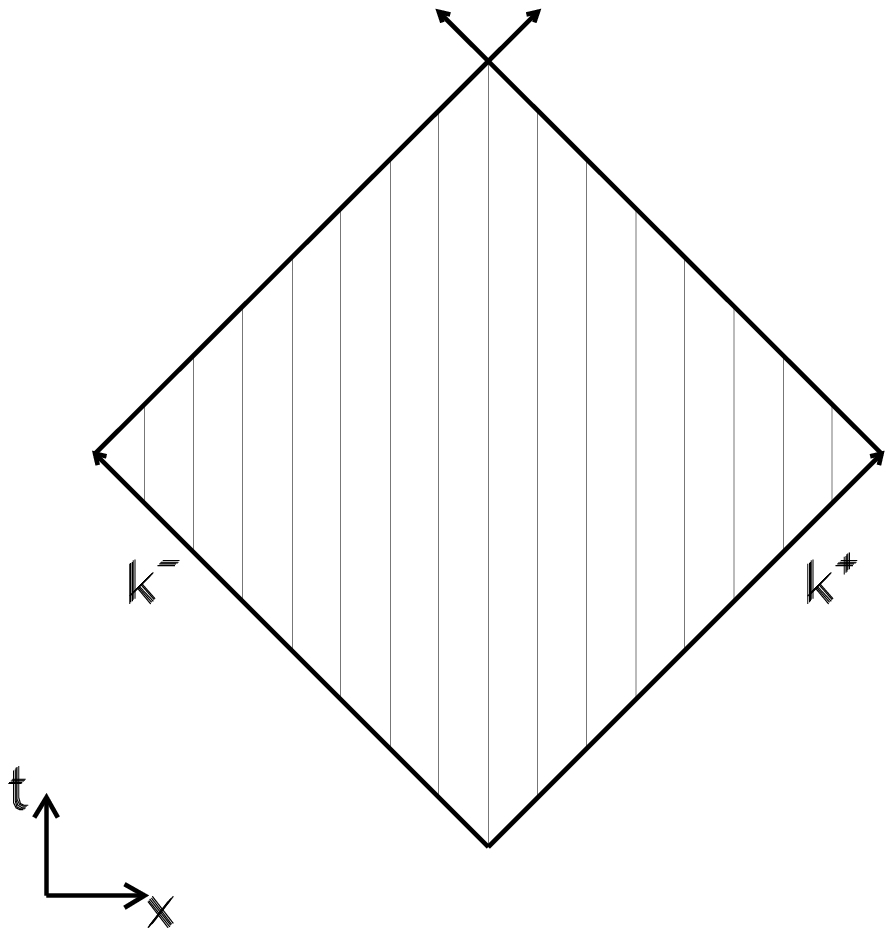,width=5.cm}}
\vspace{-0.5cm}
 \caption{\protect\small \sl
 On the left: the phase of the modified helix
  (Eq.\ref{eq:my_helix})
 for parameter $S$=0.5 rad/GeV and invariant mass of the string
 $M_0$=91.22 GeV.
 The phase was fixed by a random choice at one point. On the right:
 for better illustration of the phase evolution in the space-time
 coordinates, the points with equal phases are connected by lines
 in equidistant phase intervales.
\label{fig:my_helix}
}}
   As shown in the string diagram of Fig.\ref{fig:my_helix}, the
 definition Eq.\ref{eq:my_helix} corresponds 
 to a helix with a constant pitch along
 the string (proportional to the energy density of the string).
 The phase in the modified helix scenario is constant in time
 for a given point along the string axis, forming
 a stationary wave (similar to the interference pattern due
 to an emission from two sources).
   
\section{Parametrization of the helix string: phenomenology}

 In this section we turn our attention to observable effects related to the helix ordered
 string. We shall first study a simple quark-antiquark system to get a better understanding
 of the differences between models.   

  In the Lund helix model \cite{lund_helixm}, the phase difference is directly related
 to the rapidity difference.  The fragmentation of the Lund string
 produces roughly one hadron per unit of rapidity. 
 The hadrons, therefore, obtain -- on average --
 a transverse momentum of about the same size, i.e. roughly
 independent of $y$ , and the helix-like structure
 should be visible in their azimuthal angle difference as a function of the rapidity
 ordering.

   The observable which should reveal such a behaviour was introduced in ~\cite{lund_helixm} 
\begin{equation}
\label{eq:scr}
    Screwiness(\omega) =  \sum_e P_e |\sum_{j}\exp(i(\omega y_j - \Phi_j))|^2,
\end{equation}
  where $y_j,\Phi_j$ stand for the rapidity and the azimuthal angle
 of final hadrons, $P_e$ is a normalization factor and the parameter $\omega$ is
 the characteristic frequency of the helix rotation.
 The first sum goes over hadronic events, the second one over
 hadrons in a single event.

   The expected signal for charged final particles is shown in Fig.~\ref{fig:scr}. The presence of a Lund helix field is visible as a peak
 at $\omega \sim 1/\tau$, but the significance of the peak, with respect to standard Lund fragmentation, decreases with $\tau$. 
 There is some screwiness signal in the modified helix scenario, too, but it comes in a form of a multi-peak structure difficult
 to interpret. (It is worth remembering that the experimental study of screwiness \cite{alessandro} found a few percent
 difference between data and the standard Lund model but the signal did not exhibit a single peak shape.) 

\FIGURE[htb] {
  \mbox{\epsfig{file=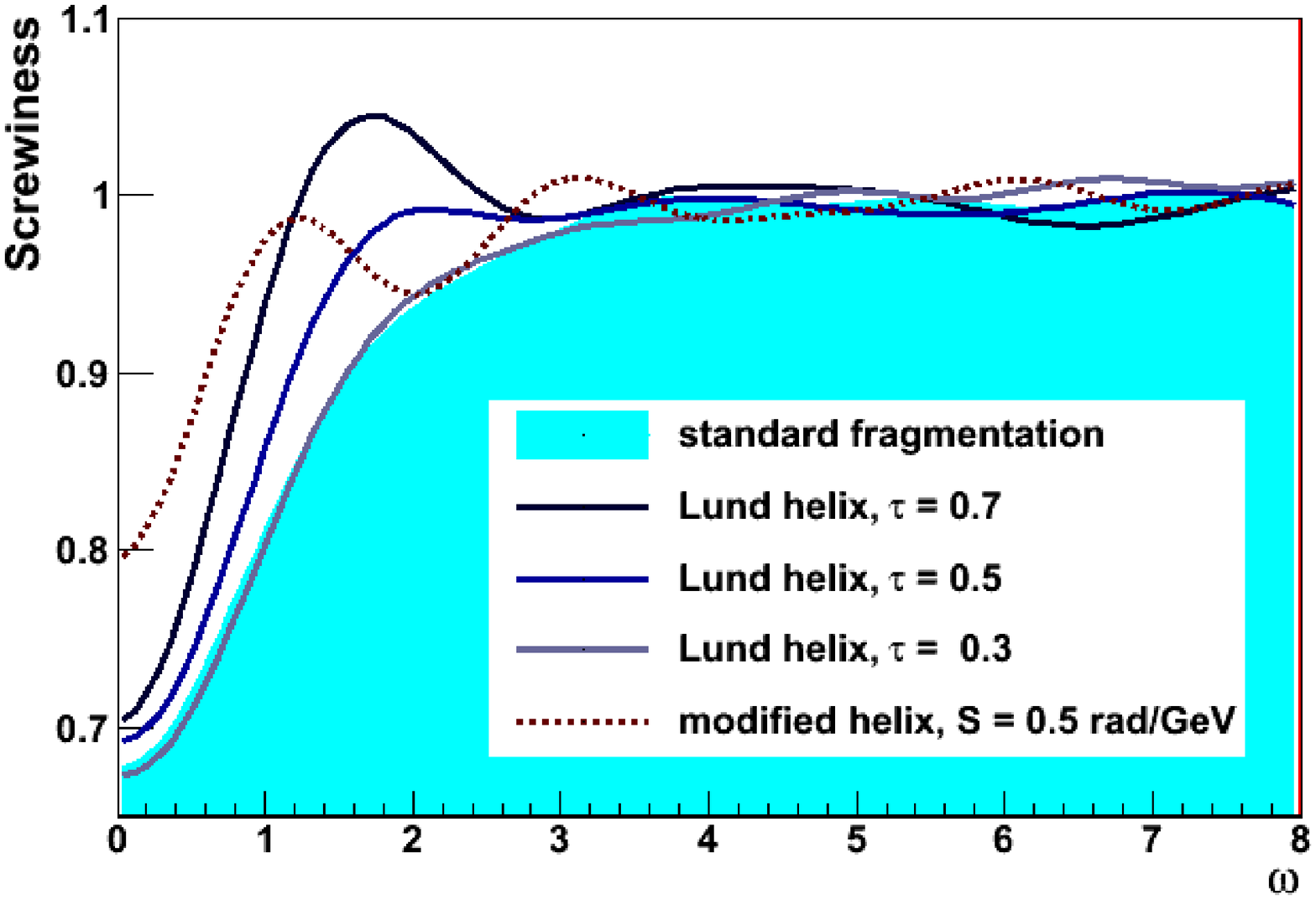,width=13.cm}}
 \caption{\protect\small \sl
  Screwiness signal obtained at the generator level in hadronic $Z^0$ decay (without parton shower),
 for various values of parameter $\tau$ in Lund helix model (3.1),
 and compared to standard Lund string fragmentation (histogram). 
 A peak is expected at $\omega \sim 1/\tau$ but its significance is small for low $\tau$ values ($ < 0.3$).
 Modified helix model (3.8) produces a multi-peak pattern (dashed line). Screwiness is calculated
 from final charged hadrons with $p \ > \ 0.15 \ GeV/c$ and $|y| < 2$.
\label{fig:scr}     
}
}

    The observable effects stemming from a modified helix parametrization are however not restricted to the hadron ordering
 in the azimuthal angle (though we will come back to the question in section 7).

   The modified helix introduces a strong correlation
 between the size of the fraction of string forming a hadron (i.e., the energy of the hadron in the string c.m.s.)
 and the size of its transverse momentum. In the rest frame of the string,
\begin{equation}
      {p}_T^2({\rm hadron}) = 4 \ r^2 \ \sin^2\frac{\Delta \Phi}{2}
                    = 4 \ r^2 \ \sin^2\frac{{\cal S} E_{had}}{2}.
\label{eq:pt}
\end{equation}
 where $r [GeV]$ (the 'radius' of the helix) is a parameter
 and $\Delta \Phi$ is the helix phase difference between the two break-ups which created the hadron.

\FIGURE[h] {
  \mbox{\epsfig{file=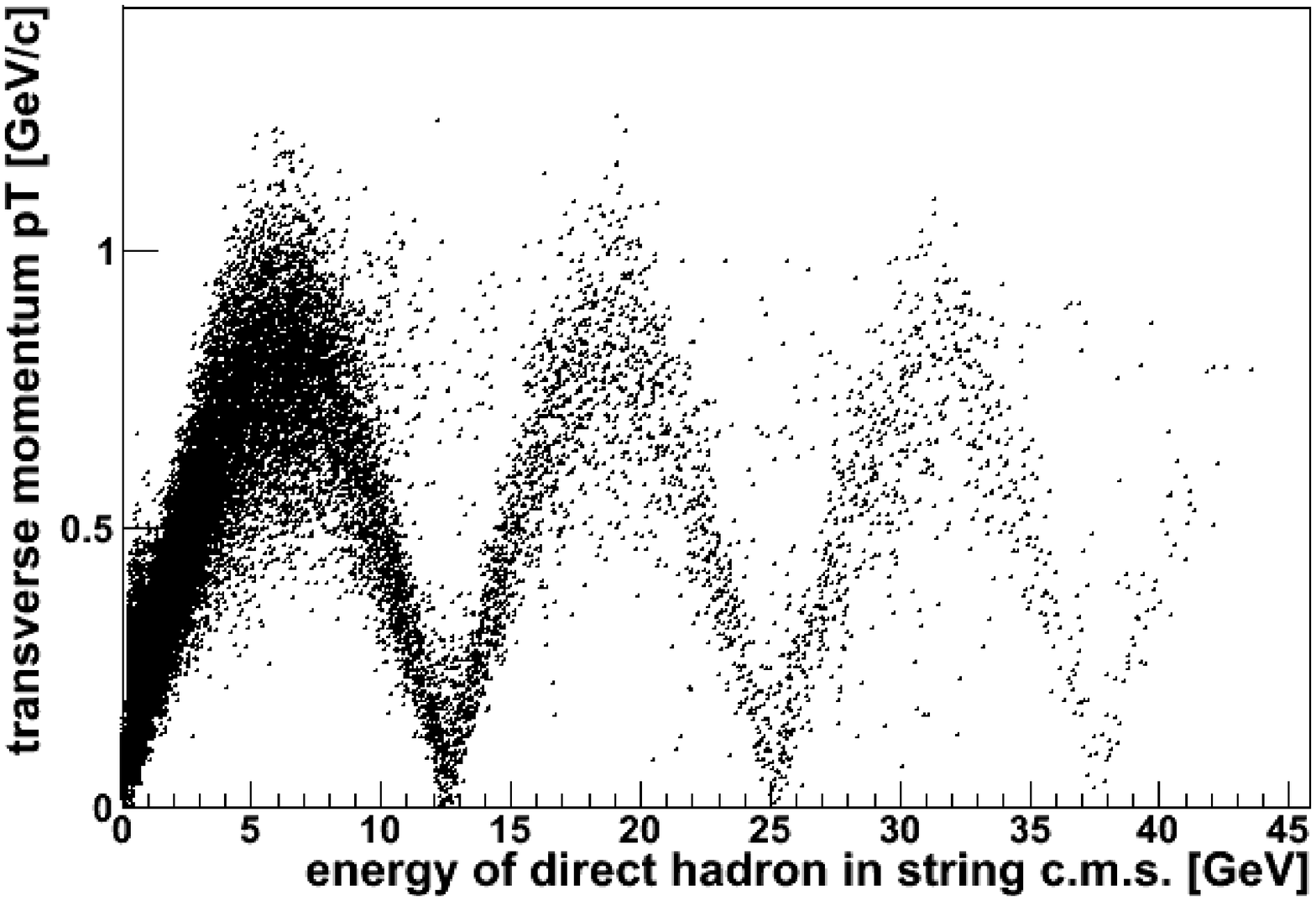,width=7cm}}
%\vspace{-0.5cm}
 \caption{\protect\small \sl
  Correlations between the transverse momentum of a direct hadron and its energy in the string c.m.s. in the
 modified helix model, r=(0.4$\pm$0.1)GeV/c, S=0.5 rad/GeV.
\label{fig:pte}     
}}

  The correlations are visible in Fig.~\ref{fig:pte}, where a clear structure appears in the distribution of direct hadrons. Experimentally, we never observe such a clear picture of string fragmentation, because it is smeared
 by the parton shower and decays. Still, these correlations leave trace in the inclusive $p_T$ spectra, as shown in Fig.~\ref{fig:ptincl}.
 Due to the exponentially falling distribution of 
 hadron energy in fragmentation which governs the size of transverse momentum through Eq.~\ref{eq:pt}, 
 the modified helix model creates more hadrons with very low $p_t$
 but less in the region $p_t \approx 0.4 GeV/c$ where the peak of the uncorrelated, gaussian distribution lies.   
\FIGURE[htb] {
  \mbox{\epsfig{file=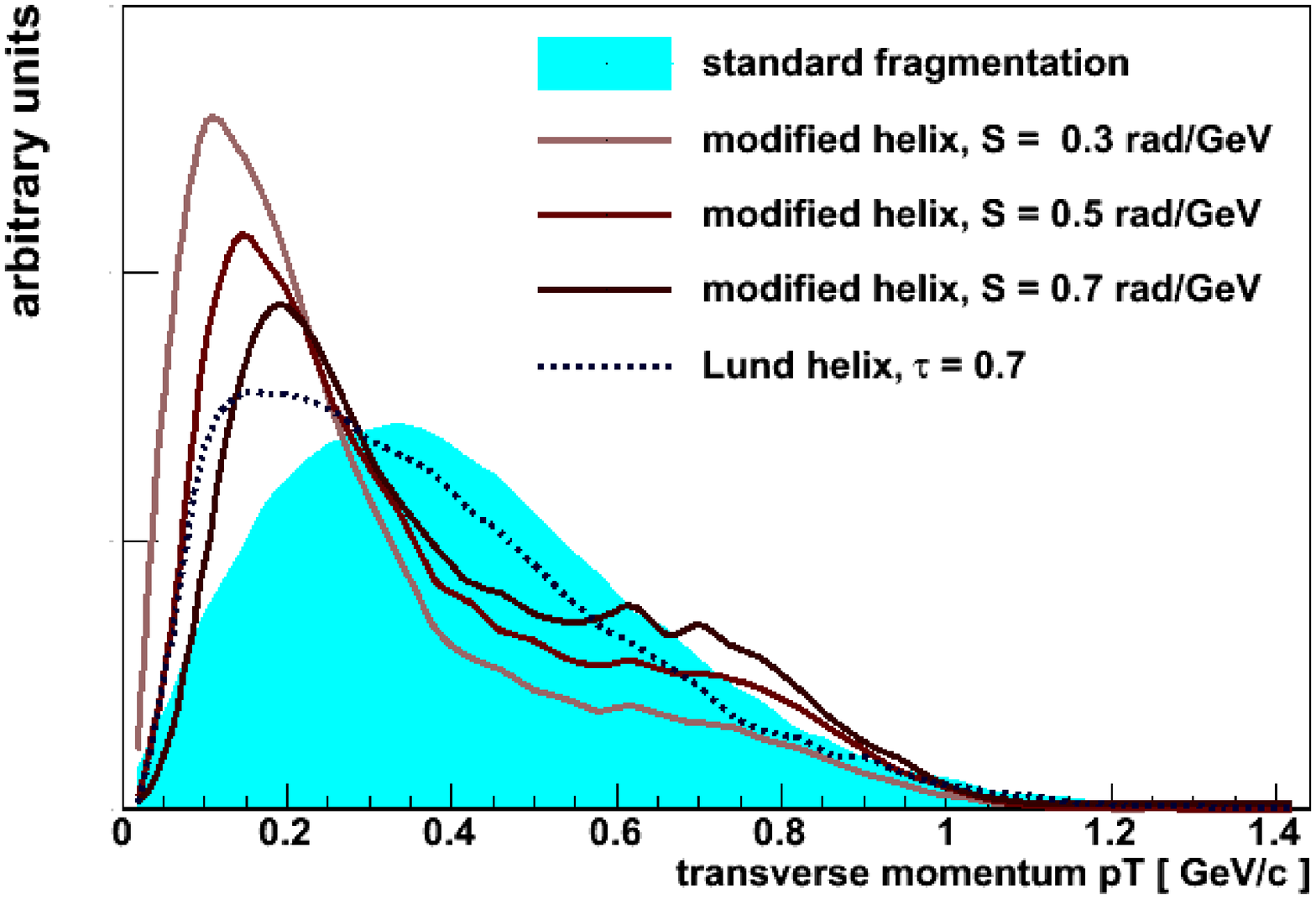,width=11cm}}
%\vspace{-0.5cm}
 \caption{\protect\small \sl
  Inclusive transverse momentum distribution of direct charged pions , for the modified helix scenario (3.8) with different helix pitch, compared to the standard fragmentation. For comparison, the distribution obtained with the Lund helix model (3.1) is shown, too (dashed line). 
\label{fig:ptincl}     
}}

\clearpage

 For comparison, Fig.~\ref{fig:ptincl} also shows the spectrum obtained from the Lund helix model with $\tau$ = 0.7,
 which exhibits qualitatively similar behaviour, albeit attenuated, as the modified helix model. (Generally speaking, we may expect
 observables designed for one helix scenario to show some effect in the other scenario, too, but weaker and  somewhat distorted, as 
 we saw in the case of the screwiness measure.)
 
  The effect of the modified helix scenario on the inclusive $p_T$ is strong enough to be readily visible in experimental data.
 As a matter of fact, a characteristic discrepancy in the low $p_T$ distribution is visible both in LEP  
 \cite{tuning} and LHC \cite{atlas} data, but before performing a direct comparison of data and the model, we need to
 make sure the model handles properly the multi-parton string we use for description of the real data.

\section{Extension of helix model on multiparton string topology}

\FIGURE[hr] {
  \mbox{\epsfig{file=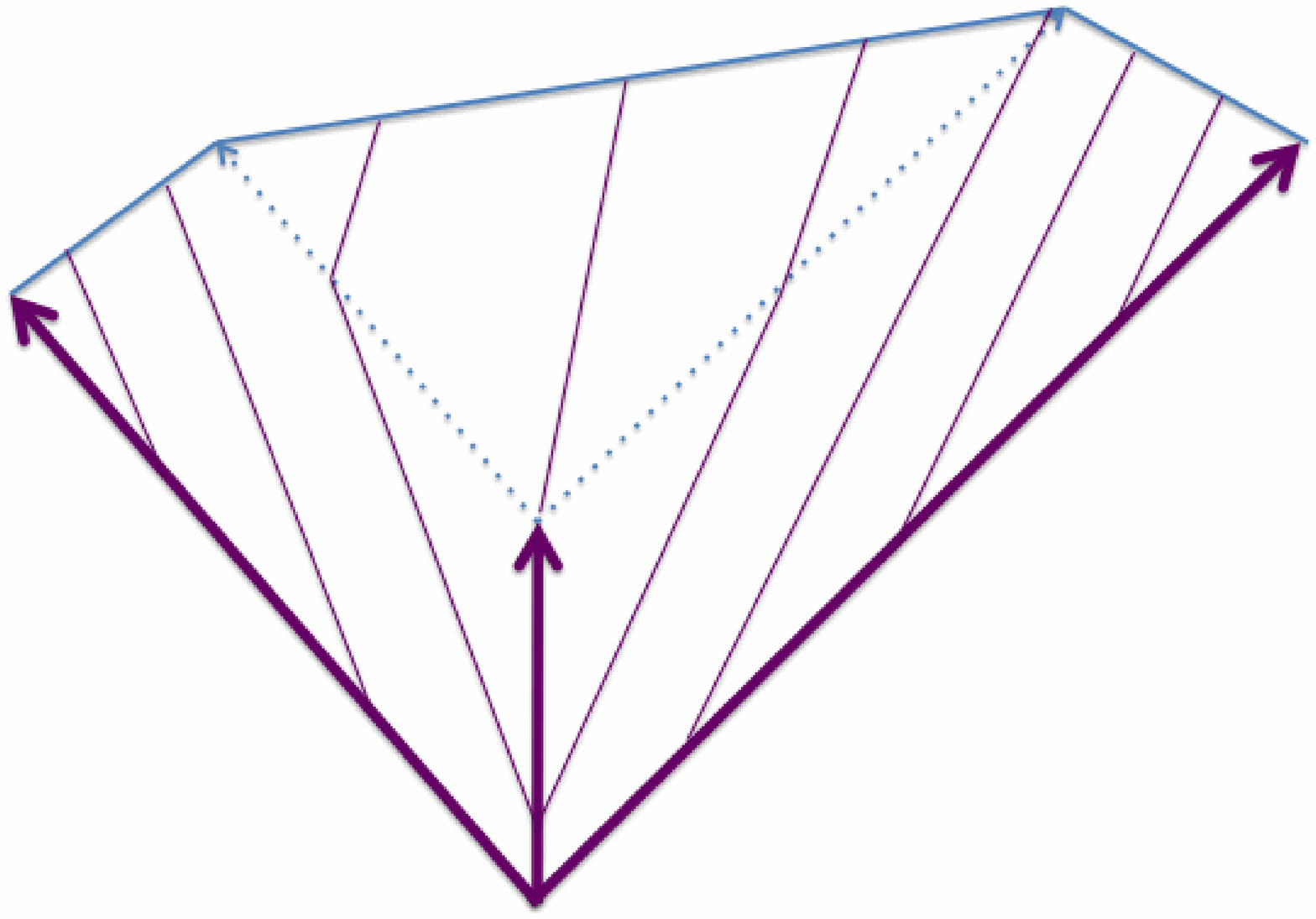,width=7.5cm}}
\vspace{-0.5cm}
 \caption{\protect\small \sl
  An illustration of helix phase evolution of the modified helix model in case of presence of a hard gluon kink on the string.
\label{fig:kink}
}}
  The comparison of the helix string model with data requires the model to be extended to cover
 not only the simple case of $q\bar{q}$ system but also an arbitrary multiparton configuration corresponding to the emission
 of hard gluons from the quark-antiquark dipole. This is actually the most complicated part of the model implementation which
 requires some additional  assumptions to be made.

  The solution adopted for the modified helix scenario  
 consists of two steps: first, the multiparton system is followed in space-time (every parton looses about 1 GeV
 of its energy per fermi in favour of the developing string field, the energy loss of gluon is twice as much because
 the gluon participates in the creation of two adjacent string pieces) in order to find the way the string breaks into
 pieces, and to evaluate their respective masses (Fig.\ref{fig:kink}). Every string piece is formed by a combination of the fractions of momenta
 of 2 partons, combinations and fractions depend on the distribution of partons in the phase space. The second step
 consists in calculation of the combined helix phase difference between endpoint quarks
\begin{equation}
  \Delta\Phi = S \sum_{i} M_{i},
\label{eq:combphase}   
\end{equation}
   where the sum runs over all (ordered) string pieces and $M_i$ is the mass of the i-th string piece. Since the phase of the modified helix string is constant at a given point along the string, it can be easily calculated from the relative distance from string endpoints. A convenient way of doing this is to use the energy fraction. For a given string break-up, for example, one can calculate the total energy
of hadrons on the left and right side. The sum of hadron energies being identical to the sum of energies of ordered string pieces, it is possible to associate the string break-up with a definite string piece, and to deduce the helix phase at a point
of string break-up using Eq.\ref{eq:my_helix}. The possibility to recover the helix phase anywhere in the string diagram in a simple way is due to the static nature of the helix field and cannot be applied to a non-static helix definition, in particular, to the Lund helix model.
     
  It should be mentioned that the modified helix model extension for hard gluon kinks uses the assumption that the helix phase
 runs smoothly over the gluon kink, i.e. the helix phases at the connecting ends of adjacent string pieces coincide.

  As much as we would like to perform a similar extension for hard gluon kinks using the Lund helix definition, we must admit the task
 goes beyond the scope of this paper. To start with, the Lund helix model corresponds to non-static helix form which evolves along the string with time. The evolution of the helix phase for a complex string topology was not addressed by the authors of the model
in \cite{lund_helixm}. Second, there is the double discontinuity in the definition of helix winding at the gluon kink which
needs to be stabilized somehow, and such a decision clearly belongs to the authors of the model.

  Further details about implementation of the modified helix string scenario in Pythia code are given in Appendix A.

 %  In this context we may argue that the modified helix model is the simplest variant of possible helix string implementations.
 %A deviation from the static concept of the helix field implies time dependence of the helix form which may become rather difficult to follow
 % in case of complicated string topologies. This argument obviously cannot be held against the Lund helix model but 
 %it has practical consequences, for example we are at the moment left with only one complete Monte-Carlo implementation
 % of a helix string, the modified helix scenario, which can be compared with the data.  

\section{Model tuning and comparison with data}
  
   The helix string model (variant \ref{eq:my_helix}) has been tuned \cite{helix_tune} to the DELPHI data \cite{tuning} using
 a set of 6 simultaneously optimized parameters: helix radius $r$ and pitch $S$, Lund fragmentation parameters $a$ and $b$,
 effective coupling constant $\Lambda_{QCD}$ and parton shower cut-off. It was known from previous studies that the helix string model
 significantly improves the description of $p_T$ spectra, but the positive impact is actually much broader. 
  
   In Table \ref{table:chi2} we see that the helix string model reduces the average $\chi^2/N_{bin}$ by more than one unit for the set of
 inclusive charged particle distributions and event shape variables used in the tune ( 619 data bins were used in total ), for both
 types of parton shower used in the study (Pythia p$_T$ ordered parton shower and Ariadne \cite{ariadne} parton shower). This is a remarkable result
 if one takes into account that the model actually removes a degree of freedom from the fragmentation process. While the performance 
 of the (modified) helix
 string model in the description of data does not amount to a proof of the existence of the helix string structure in nature, it 
 is a powerful indication that azimuthal ordering plays a significant role at the soft end of the parton cascade. 

\TABLE[b] {
%\begin{center}
\begin{tabular}{| l || c | c || c | c |} 
\hline
  Data set  &  Pythia  & helix + Pythia & Ariadne & helix + Ariadne \\  
\hline
 inclusive spectra   &  & & &  \\
 + event shapes & 4075 & 2453 & 2453 & 1489 \\
 $N_{bin}=619$  &  & & &  \\
\hline
% ident.part.rates & & & & \\
% + b-fragmentation & 444 & 669(*) & 614(*) & 586(*) \\
% $N_{bin}=47$  &  & & &  \\ 
%\hline
\end{tabular}
\caption{ The $\chi^2$ difference between the DELPHI $Z^0$ data and models, summed over inclusive charged particle spectra and event shape variables \cite{helix_tune} . The 'Pythia/Ariadne' labels distinguish
 between simulation setup using Pythia 6.421, resp. Ariadne 4.12 parton shower.
}
\label{table:chi2} 
%\end{center}
}

\FIGURE[ht] {
a)\hspace{-0.2cm}\mbox{\epsfig{file=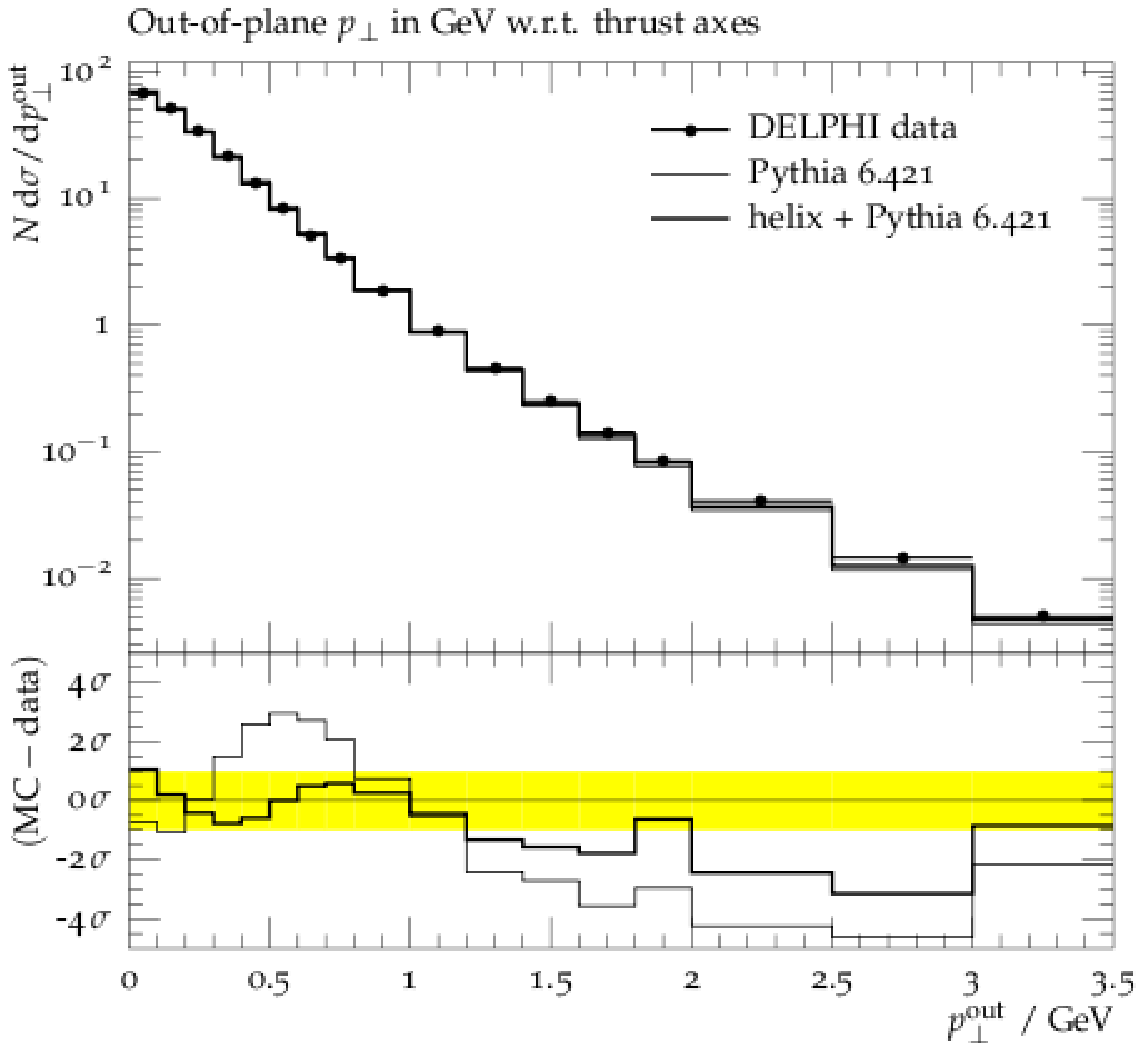,width=7.cm}}
b)\hspace{-0.2cm}\mbox{\epsfig{file=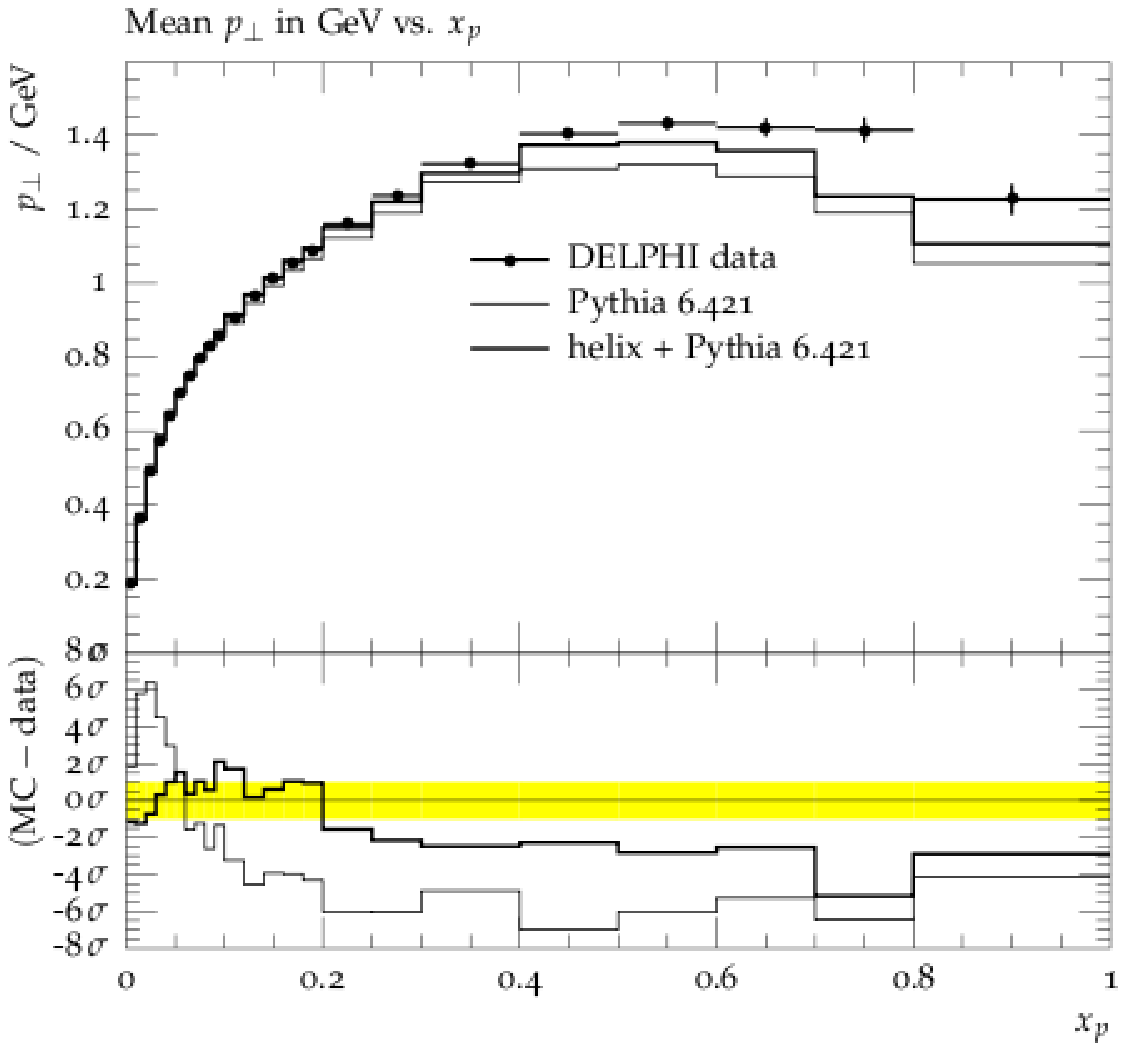,width=7.cm}}
 \caption{\protect\small \sl
  Comparison of inclusive charged particle distribution measured by the DELPHI Coll.\cite{tuning}
 and Pythia 6.421 modelling: a) $p_T$ projection on the Thrust-minor axis, b) average $p_T$ as function of scaled momentum $x_p$.
 % The results obtained with the standard fragmentation are indicated by a thin line, thick line describes the modified helix string scenario.
\label{fig:tune_pt}
}}
   The impact of the helix string model on the modelling is best seen in the comparison with measured charged $p_T$
  distributions. Figure ~\ref{fig:tune_pt} a) shows the projection of the transverse momentum of charged particles
  on the Thrust-minor axis ($p_T^{out}$). Data are compared with Professor tune of Pythia 6.421 \cite{prof_tune}
  and with the tuned helix string model, implemented in a private version of Pythia 6.421 \cite{pystrf}. The characteristic
  discrepancy around $p_T\sim$ 0.5 GeV is removed by the helix string model, and the smoother description of the low $p_T$
  region leads to a better adjustment of the parameters of the parton shower, as we can deduce
  from the improved description of the tail of the distribution. 

    Of special interest for our study is the dependence of the average $p_T$ on the size of the particle momentum, which
  should be sensitive to the type of correlations pictured in Fig.~\ref{fig:pte}. As shown   
  in Figure \ref{fig:tune_pt} b), this distribution is much better described by the helix string model than by the
  standard string fragmentation. On the basis of this particular observation and of the global results
  of the helix model tuning, we can conclude that the helix string model is favoured by the data. 
    
    Tuned helix parameters suggest a helix radius $\sim$ 0.4 GeV/c (rather well constrained) and a helix winding S $\sim$ 0.7 rad/GeV
  (with large uncertainty). The variance of the helix radius was set to 0.1 GeV/c, for simplicity. The tune did not attempt to
   resolve a possible flavour dependence of the model.

\section{Other observables} \label{sect:otherobs}

    Further experimental input for the modified helix scenario (3.8) can be expected from a study of the ordering of hadrons
  in the azimuthal angle. 
  Ideally, if we would be able to order hadrons along the string according to the fragmentation chain, the azimuthal opening
  angle for any given pair of direct hadrons should be correlated with the sum of energy of hadrons laying in between
  (the helix pitch is propotional to the string energy density).

    In analogy with Eq.\ref{eq:scr}, we define
 \begin{equation}
\label{eq:scre}
      S_E(\omega) =  P \sum_{event} |\sum_{j}\exp(i(\omega \sum_{k=0}^{j}E_k - \Phi_j))|^2,  
\end{equation}

     where $\Phi_j$ stands for azimuthal angle of a hadron, $\omega$ is a parameter, $P$ is a normalization factor. The outer sum runs over events, and the inner sum over the hadrons ordered in rapidity, resp. longitudinal momentum, in a given event.
  Such an ordering is only approximative,
  but generator level studies suggest the updated screwiness measure is sufficiently sensitive to
  provide a signal in the presence of a helix-like ordering. Under a rather stringent selection of event topologies
  (in order to suppress smearing due to the parton shower),  the signal should be visible as a peak at $\omega \sim S$,
  the parameter describing the density of helix winding in Eq.\ref{eq:my_helix}. In practice, the selection can be done
  by rejection of events where the maximal hadron $p_T$ (with respect to the Thrust axis) exceeds 1 GeV/c.

    Fig.\ref{fig:scre} shows the result of a generator level study of the $Z^0$ hadronic decay, where $S_E(\omega)$ 
  is calculated using final charged particles only. The study was done on $\sim$ 50k events per sample, retained after
  the $p_T < 1 $ GeV/c cut (original samples contained 500k events each). A few percent signal appears in the helix string
  model in comparison with the standard string fragmentation.  

%    In principle, the helix string model generates genuine hadron-hadron correlations, especially between adjacent direct hadrons.
%  The size of the combinatoric background is however substantial, and direct observation of such correlations experimentally
%  challenging. 

    There is also a possibility that the helix string ``memory'' is partially conserved in the decay of short lived
  resonancies, which would lead to an enhanced signal,  but no study has yet been done to estimate the possible effect.

 \FIGURE[htb] {
  \mbox{\epsfig{file=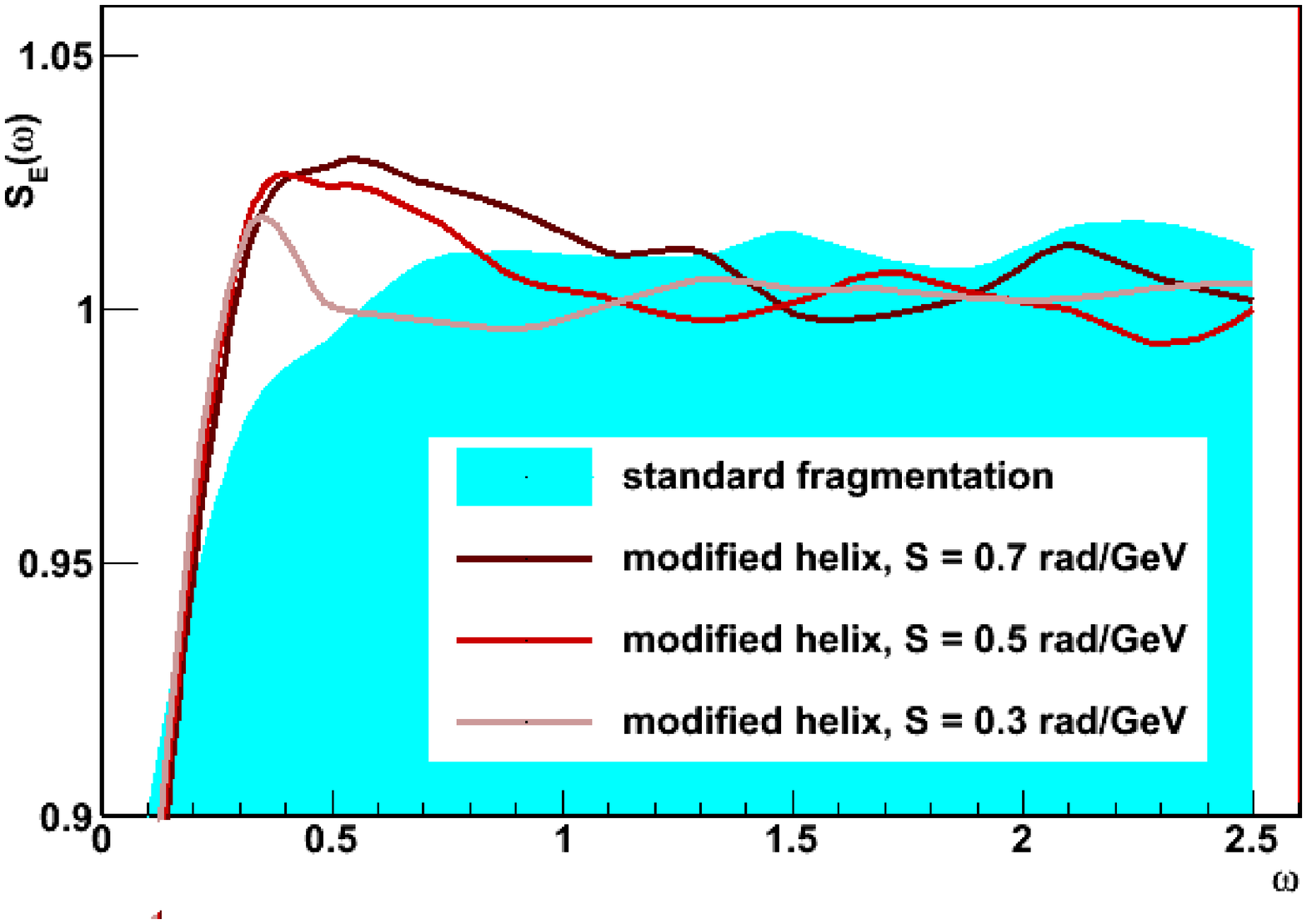,width=12cm}}
%\vspace{-0.5cm}
 \caption{\protect\small \sl
  Updated screwiness measure (Eq.\ref{eq:scre}) signal in the modified helix scenario. Generator level study with Pythia 6.421 using
 charged final particles from the hadronic $Z^0$ decay.  
\label{fig:scre}     
}}

\section{Conclusions}

   The idea of helix-like ordered gluon field definitely deserves a special attention. Further experimental evidence is needed
 to firmly establish the existence of such a phenomenon, but we are clearly touching a sensitive point in the modelling of soft
 QCD interactions. If we assume the helix ordering is regular, with pitch proportional to the distance along the string (for 
 homogenous string field), the improvement of the description of the $Z^0$ data which can be achieved with such a model
 is significant. 

   For the moment, we don't have enough information to decide for, or against, 
 a particular helix scenario, because the interplay between parton shower and formation of helix string needs further
 clarification. Hopefully, the indirect evidence gathered in this paper will contribute to a revival of interest
 for the model, in all existing variants.

%\subsection{Acknowledgments}

\bigskip

\acknowledgments

     The author would like to thank Prof.G.Gustafson for the discussion allowing to clarify 
 differences between various helix concepts. The author is grateful to the authors of the helix string model
 for providing the JETSET code including the Lund helix model modification. The tuning of the modified helix scenario was greatly simplified thanks to tools provided by Professor/Rivet projects \cite{prof}.

\appendix

\section{Pythia implementation of the modified helix scenario}
 
   The particularities of the helix string fragmentation have been discussed in section 6.2 of \cite{lund_helixm}.
 In the modified helix model, as in the original Lund helix proposal,  the transverse momentum of
 the newly created hadron is determined by the sampling of the longitudinal momentum. The algorithm
 adopted in \cite{pystrf} approximates the fragmentation function (Eq.6.8 in \cite{lund_helixm}) by

 \begin{equation}    
    f(z) = N \frac{(1-z)^a}{z} \exp{(-\frac{b}{z}(m_h^2+p_T^2(z)))}\rightarrow N \frac{(1-z)^a}{z} \exp{(-\frac{b}{z}(m_h^2+r^2))}
\end{equation}
 where $r$ is the modified helix radius, which means hadrons are initially sampled as having the transverse momentum $| p_T | = r$. Once the hadron
 momentum is calculated in this approximation, its energy is preserved while the transverse and longitudinal components are adjusted to follow
 the shape of the helix field. In case there is no kinematical solution, the $z$ fraction is resampled with the new $p_T$ estimate.

   The modified helix model is implemented in Pythia 6.421 \cite{Sjostrand:2006za} via private version of  the fragmentation routine
 PYSTRF. The following Pythia parameters and switches are used for steering:
   
\begin{itemize}
\item{MSTU(199)=} 0/2 (standard fragmentation/modified helix fragmentation)
\item{PARJ(102)=} * ( helix radius r [GeV/c], replaces PARJ(21) )
\item{PARJ(103)=} * ( helix radius variance [GeV/c] )
\item{PARJ(104)=} * ( parameter S [rad/GeV] )
\item{PARJ(109)=} 0.001 (azimuthal angle tolerance in the iterative search of the string break-up solution
 conform to Lund fragmentation rules \emph{and} helix string parametrization)  
\end{itemize}
 
  The modified helix fragmentation algorithm is about 10-20\% slower than the standard Pythia fragmentation, with failure rate well below a per mille
 level.

\end{document}